\documentclass[aps,prd,10pt,amsmath,floats,floatfix, twocolumn,superscriptaddress,nofootinbib,showpacs,longbibliography]{revtex4-2}

\usepackage{amsmath, amssymb, graphicx}
\usepackage[normalem]{ulem}
\usepackage[pagebackref=false, colorlinks=true]{hyperref}
\hypersetup{colorlinks,linkcolor=blue,urlcolor=blue}
\usepackage{mathrsfs}
\begin{document}
\preprint{}

\title{Supertranslations at Spatial and Timelike Infinities in the First-Order Formalism}
\author{Divyesh N. Solanki}
\email{divyeshsolanki98@gmail.com}
\affiliation{Indian Institute of Information Technology Allahabad (IIITA), Devghat, Jhalwa, Prayagraj-211015, Uttar Pradesh, India.}

\author{Srijit Bhattacharjee}
\email{srijuster@gmail.com}
\affiliation{Indian Institute of Information Technology Allahabad (IIITA), Devghat, Jhalwa, Prayagraj-211015, Uttar Pradesh, India.}

\date{\today}

\begin{abstract}
We study supertranslations at spatial and future timelike infinity in the first-order formalism. We relax the Ashtekar-Engle-Sloan boundary conditions to allow supertranslations at the spatial infinity and obtain the precise form of the tetrads and Lorentz connections. Employing the covariant phase space technique we obtain the Hamiltonian charges for the supertranslations at both timelike and spatial infinities. The charges obtained are shown to match the expressions reported adopting the metric-based approach.

\end{abstract}

\maketitle

\section{Introduction}
The asymptotic properties of fields and spacetimes have been studied for a long time and a myriad of interesting results are obtained. In General Relativity (GR) the study of isolated systems is often useful where the gravitational field vanishes sufficiently fast at large distances from the sources. The asymptotically flat (AF) spacetimes in GR capture such systems as the geometry approaches Minkowskian at the far region. For AF spacetimes, the conformal mapping approach advocated by Penrose indicates five specific asymptotic boundaries \cite{Penrose_1963}.  The trajectories for the massive bodies have their asymptotic future and past endpoints called the future and past timelike infinities $(i^{\pm})$. The spacelike trajectories on the other hand reach an endpoint termed as spatial infinity $i^0$. The radiations which follow null geodesics can reach (or emerge) future (from past) null infinities $\mathscr{I}^{\pm}.$ 

The question of suitable boundary conditions for the gravitational fields near the asymptotic regions is also intimately tied to the studies of asymptotic symmetries of AF spacetimes. The boundary conditions should be chosen judiciously to allow interesting physics in the asymptotic regions without hampering the desired structures. One of the most intriguing outcomes of such analyses is the discovery of the infinite dimensional Bondi-Metzner-Sachs (BMS) group as the asymptotic symmetry group that preserves the boundary conditions of the metric near the null infinities $\mathscr{I}^{\pm}$ \cite{Bondi, Bondi_Burg_Metzner, Sachs_1962}. BMS group encompasses the Lorentz symmetries and infinitely many additional symmetries called supertranslations. Supertranslations played a pivotal role in understanding the relation between gravitational scatterings, soft theorems, and memory effects \cite{Strominger_Zhiboedov, Strominger_Prahar, Strominger_review}. Several extensions of the BMS group at null infinities have also been obtained recently \cite{Barnich_Troessaer_2010, BMS/CFT, barnich_2012, Campiglia_Laddha_2014, Fuentealba_Henneaux_Tross}.  Further developments of these analyses led to the celestial holography program \cite{Aneesh_Compere}. 

The study of asymptotic structure at the spatial infinity also has a long history \cite{ADM_1959, Geroch_1972, Regge_Teitelboim, Ashtekar_Hansen, Beig_Schmidt, Ashtekar_Romano, Compere_Dehouck}. The motivation behind exploring BMS symmetry at $i^0$ lies in the formulation of quantum theory where the states are usually defined on spacelike Cauchy slices. The BMS-supertranslation charge expressions at the past and future boundaries  ($\mathscr{I}^{+}_-$ and $\mathscr{I}^{-}_+$) of null infinities should match to get a consistent conservation law. This matching involves passing through the spacelike infinity $i^0$. Therefore, it is natural to expect the existence of non-vanishing BMS charges at spatial infinity. The BMS symmetries at spatial infinity with consistent boundary conditions have been achieved only recently \cite{Troessaert_2018, Henneaux_Tross, Henneaux_Tross_EM, Compere}. These works proposed BMS charges that are consistent with future and past null infinities. The same has also been achieved in \cite{Prabhu_2019, Prabhu_2020} following the Ashtekar-Hansen framework \cite{Ashtekar_Hansen}\footnote{For a recent account in line with \cite{Ashtekar_Hansen}, see \cite{Ashtekar_Khera}.}.

BMS symmetries and their charges at the timelike infinities have been investigated recently in \cite{Sumanta_Virmani} following the studies made for the spatial infinity in the metric-based approach suggested by Beig-Schmidt \cite{Beig_Schmidt}. See, \cite{Cutler_1989, Porrill_1982, Gen_1998} for earlier works. The purpose of searching for BMS symmetries at the timelike infinities is related to the recent developments in the gravitational scattering and soft theorems \cite{Laddha_Campiglia, Strominger_review}. There have been suggestions that the horizon of a stationary black hole should contain non-zero BMS-like charges \cite{Donnay_PRL_2015, Hawking_Perry_Strominger_2016, Hawking_Perry_Strominger_2017, Strominger_review}. The charge (supertranslation) conservation in the presence of black holes requires contribution from both the horizon and the null infinity. Thus, one needs to find a relationship between the charges at the null infinity and the horizon. In \cite{C_F_Prabhu}, a possible way to relate the charges at these boundaries is suggested by involving timelike infinities which are situated at the corners of the horizons and null infinities. For a comprehensive picture of the present status of asymptotic symmetries at the five corners in a metric-based approach, readers are suggested to see \cite{Compere}.

In this paper, we study the supertranslations at the spatial and the future timelike infinity in a first order framework where the gravitational fields are described by the tetrads and spin (Lorentz) connections. Earlier works using the first order framework chiefly focused on the Poincar\'e charges at the spatial and null boundaries and on the laws of black hole mechanics in quasi local setups in the Palatini gravity or with the extended versions of it \cite{Ashtekar_Krishnan, Ashtekar_Krishnan_living_review, Ashtekar, Corichi_2010, Corichi_Garcia_2014, Chatterjee:2020iuf}.   First order formalism has proven to be useful to compute quantities of interests in the phase space with less labor than the metric-based second-order methods. In this note, we extend the pioneering contribution by Ashtekar-Engle-Sloan \cite{Ashtekar} on the asymptotic analysis at spatial infinity. See \cite{Ashtekar_Sloan_hd} for a higher dimensional version of the study made in \cite{Ashtekar}. In \cite{Ashtekar}, a finite symplectic structure and Poincar\'e charges were obtained at spatial infinity in 4-dimensions by considering boundary conditions that ruled out supertranslations and logarithmic translations\footnote{The asymptotic symmetries at spatial infinity (and at timelike infinities) also contain logarithmic translations \cite{Ashtekar_Hansen, Ashtekar_1985, Sumanta_Virmani}. Recently, in \cite{Fuentealba_Henneaux_Tross}, a new symmetry called the logarithmic supertranslations has been found at spatial infinity.} at $i^0$.  We find a finite symplectic structure and obtain the covariant Hamiltonian charges for the supertranslations both at spatial and timelike infinities considering a set of boundary conditions inspired by the works \cite{Compere_Dehouck, Sumanta_Virmani, Troessaert_2018}. The supertranslation charges obtained in our study match with the charges reported in the literature \cite{Compere_Dehouck, Sumanta_Virmani}.

The paper is organized as follows. In section (\ref{ASSI}), we describe the AF Beig-Schmidt metric at $i^0$ and discuss the boundary conditions. In section (\ref{First-Order}), we provide a brief overview of the first-order framework and derive the tetrad and the Lorentz connection for the Beig-Schmidt metric. In section (\ref{CPS}), we define the Hilbert-Palatini action and discuss its finiteness. We then construct the covariant phase space and show that the symplectic structure is well-defined for the constructed tetrads and connections. Next, using the Hamiltonian vector field for supertranslation we derive its charges at $i^0$. We compare our results with the charges obtained using a metric-based approach. In section (\ref{ASFTI}), we follow a similar analysis at $i^+$ and propose the supertranslation charges. These charges also reduce to the known results. Finally, we conclude by summarizing our work and providing some future prospects in section (\ref{Discussions}). 


\section{Asymptotic Structure at Spatial Infinity $i^0$}
\label{ASSI}
We begin this section by defining a suitable coordinate system to perform the asymptotic analysis at $i^0$. We follow the  Beig-Schmidt \cite{Beig_Schmidt} analysis to describe the asymptotic metric at $i^0$. 


\subsection{Flat spacetime at $i^0$}
Consider the four-dimensional Minkowski metric $\eta$. If $x^a$ are the Cartesian coordinates for $\eta$, then outside the lightcone at any point $p$ we have $\eta_{\mu\nu}x^\mu x^\nu > 0$. Thus, we define $\rho=\sqrt{\eta_{\mu\nu}x^\mu x^\nu}=\sqrt{r^2-t^2}$ as a radial coordinate, where $r^2=(x^1)^2+(x^2)^2+(x^3)^2$. Now, the timelike coordinate $\tau$ is defined as
\begin{equation}
    \tau=\tanh^{-1}(t/r)
\end{equation}
whose ranges are chosen to be $(-\infty,\infty)$ so that it allows both incoming and outgoing trajectories. Thus, $i^0$ can relate $\mathscr{I}^+$ and $\mathscr{I}^-$ which include outgoing and incoming trajectories respectively.
Using the following transformations
\begin{align}
    &r=\rho\cosh{\tau},
    &t=\rho\sinh{\tau},
\end{align}
the Minkowski metric in $(\tau,\rho,\theta,\phi)$ coordinates can be written as
\begin{equation}
    ds^2 = d\rho^2 + \rho^2 \big(-d\tau^2 + \cosh^2{\tau} \  \gamma_{AB} dx^A dx^B \big).
    \label{Min_i0}
\end{equation}
Where $\gamma_{AB}$ is a metric on the unit sphere.

We rewrite (\ref{Min_i0}) in the compact form:
\begin{equation}
    ds^2 = d\rho^2 + \rho^2 h^{(0)}_{ij}d\Phi^i d\Phi^j,
\end{equation}
where the coordinates $\Phi^i=(\tau,\theta,\phi)$ parametrize $\rho=\text{constant}$ Hyperboloids $\mathcal{H}$, $h^{(0)}_{ij}$ is the induced metric on the unit Hyperboloid $\mathcal{H}$. In these coordinates, $i^0$ is approached as $\rho\to\infty$ leaving other coordinates $\Phi^i$ finite.


\subsection{Asymptotically flat spacetime at $i^0$}
\label{AFSTi0}
As we know we cannot find a unique Minkowski metric $\eta_{ab}$ at $i^0$, because there exist non-rigid coordinate freedoms (eg. supertranslations) that allow us to recover $\eta_{ab}$ only up to order $\mathcal{O}(\rho^{-1})$. To write down such a metric let us first define the asymptotic expansion of any function $f$ of interest at $i^0$:
\begin{equation}
    f(\rho,\Phi) = \sum_{n=0}^{m} \dfrac{f^{(n)}(\Phi)}{\rho^n} + o(\rho^{-m}).
\end{equation}
Where the small ``$o$" notation indicates that the remainder can also include logarithmic terms. It has a property that $$\lim_{\rho\to\infty} \rho^m o(\rho^{-m})=0.$$
The asymptotic flatness is defined as follows. A smooth spacetime metric $g$ outside the lightcone of some point $p$ is said to be weakly asymptotically flat at spatial infinity if there exists a Minkowski metric $\eta$ such that outside the spatially compact worldtube $g-\eta$ has a fall-off of order one  \cite{Ashtekar}, i.e., $$\lim_{\rho\to\infty} (g-\eta)=0.$$
Thus the asymptotically flat spacetime at $i^0$ reduces to eq. (\ref{Min_i0}) as $\rho\to\infty$.
This implies that in the $(\rho,\Phi)$ coordinates associated with $\eta$ the spacetime metric $g$ can be asymptotically expanded in the most general form as follows:
\begin{multline}
    ds^2 = \bigg(1+\dfrac{2\sigma}{\rho} + \dfrac{\sigma^2}{\rho^2} + o(\rho^{-2}) \bigg) d\rho^2 + o(\rho^{-2})_i\rho d\rho d\Phi^i\\
    + \rho^2 \bigg(h^{(0)}_{ij} + \dfrac{h^{(1)}_{ij}}{\rho} + \dfrac{\log{\rho}}{\rho^2} i_{ij} + \dfrac{j_{ij}}{\rho^2} + o(\rho^{-2})_{ij} \bigg) d\Phi^i d\Phi^j,
    \label{g_general}
\end{multline}
where $\sigma$ and $h^{(1)}_{ij}$ are the first-order tensor fields on $\mathcal{H}$; $i_{ij}$ and $j_{ij}$ are the second-order tensor fields on $\mathcal{H}$. The log term, i.e., $\log{\rho}$, is counted as order unity. From the vacuum Einstein field equation one can show that $\sigma$ satisfies 
\begin{equation}
    (D^2+3)\sigma=0,
    \label{sigma_diff}
\end{equation}
where $D^2=D^iD_i$ is the Laplacian on $\mathcal{H}$. The general solution of (\ref{sigma_diff}) contains even and odd parity harmonics. The metric (\ref{g_general}) has a coordinate freedom to perform the logarithmic translations and the supertranslations. Thus, the asymptotic symmetry group turns out to be larger than the BMS group at $i^0$. In the present study, we adopt appropriate boundary conditions to eliminate the logarithmic translations and recover the BMS group at $i^0$ \cite{Compere_Dehouck, Troessaert_2018}.

Following \cite{Compere}, let us define the tensor potential $k_{ij}$ on $\mathcal{H}$ as $k_{ij}=h^{(1)}_{ij}+2\sigma h^{(0)}_{ij}$. The coordinate freedom allows us to set:
\begin{align}
    &k = h^{(0)ij}k_{ij} = 0, \label{BC1}\\
    &D^ik_{ij}=0. \label{BC2}
\end{align}
Now, we assume that $k_{ij}$ is defined by a scalar smooth function $\Psi$ on $\mathcal{H}$:
\begin{equation}
    k_{ij} = -2(D_iD_j + h^{(0)}_{ij})\Psi,
\end{equation}
which is equivalent to the vanishing of the leading component of the magnetic part of the Weyl tensor. The boundary conditions on $k_{ij}$ imply that $\Psi$ satisfies
\begin{equation}
    (D^2+3)\Psi = 0.
\end{equation}
The arbitrary function $\Psi$ also satisfies the same differential equation (\ref{sigma_diff}), whose general solution can be written as a sum of even- and odd-parity harmonics. Considering $\sigma$ to be an even-parity function and $\Psi$ to be an odd-parity function eliminates the logarithmic translations but retains the supertranslations. This fact was elucidated by Troessaert in \cite{Troessaert_2018} using the boundary conditions proposed by Compère-Dehouck \cite{Compere_Dehouck}. Throughout this paper, we assume that $\sigma$ is an even-parity function on $\mathcal{H}.$\footnote{For the details on the asymptotic analyses at $i^0,\ i^{\pm}$, and the boundary conditions readers are suggested to see \cite{Compere}.}

The Beig-Schmidt expansion up to the first-order fields in eq. (\ref{g_general}) is sufficient to study the supertranslations at $i^0$. Thus, we truncate the metric up to the first order fields, and get:
\begin{multline}
    ds^2 = \bigg(1+\dfrac{2\sigma(\Phi)}{\rho} \bigg) d\rho^2 + \bigg(h^{(0)}_{ij}+\dfrac{h^{(1)}_{ij}}{\rho} \bigg) (\rho d\Phi^i) (\rho d\Phi^j) \\+ o(\rho^{-1}).
    \label{Beig_Schmidt}
\end{multline}
The diffeomorphism
\begin{align}
    &\rho\to\rho + \omega(\Phi) +o(\rho^0),\\
    &\Phi^i\to\Phi^i + \rho^{-1}D^i\omega + o(\rho^{-1}),
\end{align}
preserves the above form of the metric, where $\omega$ is an arbitrary function of the angular coordinates $\Phi$. The above diffeomorphism is a general class of supertranslations at $i^0$. The function $\sigma$ and the field $k_{ij}$ transform as
\begin{align}
    &\sigma \to \sigma, \\
    &k_{ij} \to k_{ij} +2 D_i D_j \omega + 2\omega h^{(0)}_{ij}.
\end{align}
The supertranslations that preserve the boundary conditions (eqs. (\ref{BC1}), (\ref{BC2})) must satisfy
\begin{equation}
    \left(D^2 + 3\right)\omega = 0.
\end{equation}
This is the specific class of supertranslations at $i^0$ for which we will derive the supertranslation charges.


\section{First-order framework}
\label{First-Order}
In the first-order formalism, the metric tensor and the connection are considered independent dynamical fields. Consider an isomorphism between the tangent space at a point $p$ on the manifold ($T_pM$) and the internal vector space ($V$) equipped with the Minkowski metric $\eta_{IJ}$:
\begin{equation}
    e:T_pM \to V.
\end{equation}
It has the components $e_a^I$, where $a,b,c,...$ are spacetime indices and $I,J,K,...$ are internal indices.
The covariant derivative on the internal space compatible with the spacetime metric $g_{ab}$ and the Minkowski metric $\eta_{IJ}$ is defined as
\begin{equation}
    \mathcal{D}_a u_b^{I} = \nabla_a u_b^I + \omega_{a \ J}^{\ I} u_b^J,
    \label{D_au_b^I}
\end{equation}
where $u_b^I$ is any one-form, $\omega_{a}^{\ IJ}$ is a connection one-form\footnote{This $\omega_a^{IJ}$ should not to be confused with the supertranslation parameter $\omega(\Phi)$.}, and $\nabla_a$ is a covariant derivative operator on the spacetime. Since $\mathcal{D}_a\eta_{IJ}=0$, the connection one-form is antisymmetric in the internal indices, i.e., $\omega
_{aIJ}=-\omega_{aJI}$.\\
In form notations, the above equation is written as
\begin{equation}
    \boldsymbol{\mathcal{D}u}^I = d\boldsymbol{u}^I + \boldsymbol{\omega}^{I}_{\ J} \wedge \boldsymbol{u}^J.
\end{equation}
If we replace $\boldsymbol{u}^I$ with $\boldsymbol{e}^I$, we obtain the first Cartan structure equation
\begin{equation}
    \boldsymbol{\mathcal{D}e}^I=\boldsymbol{de}^I + \boldsymbol{\omega}^{I}_{\ J} \wedge \boldsymbol{e}^{J} = 0,
\end{equation}
which is also used to establish the compatibility relation between $\boldsymbol{e}$ and $\boldsymbol{\omega}$. The spacetime metric $g_{ab}$ and the Minkowski metric $\eta_{IJ}$ are related by
\begin{equation}
    g_{ab} = \eta_{IJ}e^{I}_{a} e^{J}_{b}.
    \label{gnee}
\end{equation}
We consider a fiducial derivative operator $\bar{\partial}_a$, which is torsion-free and compatible with a fixed co-frame $e_b^{(0)I}$ such that $\bar{\partial}_ae_b^{(0)I} = 0$. With this choice, the covariant derivative $\nabla_a$ in eq. (\ref{D_au_b^I}) reduces to $\bar{\partial}_a$. We will use Cartesian coordinates as well as the radial hyperboloid coordinates $(\rho,\ \Phi^i)$ to perform the asymptotic expansion of the co-tetrad about the fixed co-frame.

The asymptotic expansion of the tetrad $\boldsymbol{e}^I$ and the connection one-form $\boldsymbol{\omega}^{IJ}$ at spatial infinity is given by
\begin{equation}
    \boldsymbol{e}^I = \boldsymbol{e}^{(0)I} + \dfrac{\boldsymbol{e}^{(1)I}}{\rho} + o(\rho^{-2}),
    \label{e_expan}
\end{equation}
\begin{equation}
    \boldsymbol{\omega}^{IJ} = \boldsymbol{\omega}^{(0)IJ} + \dfrac{\boldsymbol{\omega}^{(1)IJ}}{\rho} + \dfrac{\boldsymbol{\omega}^{(2)IJ}}{\rho^2} + o(\rho^{-3}).
    \label{w_expan}
\end{equation}
Now, we substitute eq. (\ref{e_expan}) into eq. (\ref{gnee}):
\begin{equation}
    g_{ab}=\eta_{IJ} e_a^{(0)I} e_b^{(0)J} + \dfrac{2}{\rho} \eta_{IJ} e_a^{(0)I} e_b^{(1)J} + o(\rho^{-2}),
\end{equation}
where $g^{(0)}_{ab}=\eta_{IJ} e_a^{(0)I} e_b^{(0)J}$ is a flat metric. Equating the above equation with eq. (\ref{Beig_Schmidt}), we obtain
\begin{equation}
    e_a^{(1)I} = \sigma \rho_a \rho^I  + \dfrac{1}{2} h^{(1)}_{ab} e^{(0)bI}.
    \label{e^(1)}
\end{equation}
where $\rho_a=\partial_a\rho$ and $\rho^I=\rho_a e^{(0)aI}$. In terms of the field $k_{ab}$ it is written as
\begin{equation}
    e_a^{(1)I} = \sigma (2\rho_a \rho^I - e^{(0)I}_{a} ) + \dfrac{1}{2} k_{ab} e^{(0)bI}.
    \label{e(1)_in_kab}
\end{equation}
We must note that $\rho_a k^{ab} = \rho_b k^{ab} = 0$. The last term in the above expression has appeared due to the Compère-Dehouck-Troessaert boundary condition \cite{Troessaert_2018, Compere_Dehouck}; this was absent in \cite{Ashtekar}. We convert the internal derivative operator into the fiducial derivative operator by $\partial^I =e^{(0)aI}\bar{\partial_a}$ to calculate $\boldsymbol{\omega}^{IJ}$ without using the components of $\boldsymbol{e}^{(0)I}$ explicitly. The unique torsion-free connection one-form in terms of $\boldsymbol{e}^I$ is defined as \cite{supergravity}
\begin{equation}
    \omega_{a}^{\ IJ} = 2 e^{b[I}\partial_{[a} e_{b]}^{\ J]} - e^{b[I}e^{|c|J]} e_{aK} \partial_b e_{c}^{\ K}.
    \label{w}
\end{equation}
Substituting eqs. (\ref{e_expan}) and (\ref{w_expan}) into eq. (\ref{w}), we obtain
\begin{align}
    &\omega_a^{(0)IJ} = 0 = \omega_a^{(1)IJ},\\
    &\omega_a^{(2)IJ} = 2\rho^2 \partial^{[J} \big(\rho^{-1}e_a^{(1)I]} \big). \label{w(2)}
\end{align}
\begin{multline}
    \omega_a^{(2)IJ} = 4\rho\rho_a \rho^{[I}\partial^{J]}\sigma - 2\rho e^{(0)[I}_{a}\partial^{J]}\sigma - 2e^{(0)[I}_a \rho^{J]}\sigma
    \\+ \rho e^{(0)b[I} \partial^{J]}k_{ab} - k_{ab} e^{(0)b[I}\rho^{J]}.
\end{multline}
In terms of the $(\rho,\ \Phi^i)$ chart, the above expression is understood as follows. Note that every derivative of the $\sigma$ and $k_{ij}$ w.r.t. the hyperboloidal coordinates generates a $1/\rho$ factor, as  $\partial^J\sigma=e^{(0)aJ}\nabla_a\sigma = e^{(0)iJ}\frac{1}{\rho}D_i\sigma$ and $\partial^Jk_{ab}=e^{(0)cJ}\nabla_ck_{ab} = e^{(0)iJ}\frac{1}{\rho}D_ik_{ab}$. As a result the right-hand side of $ \omega_a^{(2)IJ}$  becomes independent of $\rho$. 


\section{Covariant Phase Space and Hamiltonian Charges}
\label{CPS}
We start this section with a brief discussion of the variational principle in the first-order formalism with the boundary conditions chosen in the previous section.


\subsection{Action principle}
Consider a four-dimensional manifold $M$ bounded by two spacelike hypersurfaces $\Sigma_1$ and $\Sigma_2$. We decompose the boundary of $M$ as $\partial M = \Sigma_1 \cup \Sigma_2 \cup \tau_\infty$, where $\tau_\infty$ is a timelike cylinder at infinity. The spatial infinity is approached through the time-like hyperboloids depicted in Fig. (\ref{fig1}). There can also be an inner boundary, but we are not considering it in this paper. The boundary of interest would be such that the compatibility relation between $\boldsymbol{e}$ and $\boldsymbol{\omega}$ is satisfied. We consider an internal timelike vector $n^I$ and partially fix the gauge such that $\partial_a n^I = 0$. Further, the allowed field configurations are those for which $n^a = e^a_I n^I$ is the unit normal to $\Sigma_1$ and $\Sigma_2$. Given this, the Hilbert-Palatini action with the Gibbons-Hawking \cite{Gibbons_Hawking, Hawking_Horowitz} boundary term is given by \cite{Ashtekar_book}:
\begin{equation}
    S = -\dfrac{1}{2\kappa} \int_M \boldsymbol{\Sigma}_{IJ} \wedge \boldsymbol{F}^{IJ} + \dfrac{1}{2\kappa} \int_{\partial M} \boldsymbol{\Sigma}_{IJ} \wedge \boldsymbol{\omega}^{IJ},
    \label{S_modified}
\end{equation}
where $\kappa = 8\pi G$, $\boldsymbol{\Sigma}_{IJ}=\frac{1}{2} \epsilon_{IJKL} \boldsymbol{e}^K \wedge \boldsymbol{e}^L$, and $\boldsymbol{F}^{IJ} = \boldsymbol{d\omega}^{IJ} + \boldsymbol{\omega}^I_{\ K} \wedge \boldsymbol{\omega}^{KJ}$ is the curvature two-form of $\boldsymbol{\omega}$. The above action (with the boundary term ) is differentiable. It is also gauge invariant if we only allow the gauge transformations that preserve the asymptotic conditions \cite{Corichi_2015, Corichi}. Varying the action the boundary term gets canceled, and yields
\begin{multline}
    \delta S = -\dfrac{1}{2\kappa} \int_M \big(\epsilon_{IJKL}\boldsymbol{F}^{IJ}\wedge \boldsymbol{e}^K \wedge \delta \boldsymbol{e}^L -\boldsymbol{\mathcal{D}\Sigma}_{IJ} \wedge \delta\boldsymbol{\omega}^{IJ} \big)\\
    +\dfrac{1}{2\kappa} \int_{\partial M} \delta\boldsymbol{\Sigma}_{IJ} \wedge \boldsymbol{\omega}^{IJ}.
    \label{dS2}
\end{multline}
Where, $\delta\boldsymbol{\Sigma}_{IJ} = \epsilon_{IJKL}\boldsymbol{e}^K \wedge \delta \boldsymbol{e}^L$. 

We know for the Einstein-Hilbert action, the variational principle is well posed in AF spacetimes if one subtracts an infinite counterterm $C$ with the trace of the extrinsic curvature ($K_0$) of the boundary manifold $\partial M$ embedded in AF spacetimes \cite{Gibbons_Hawking, Hawking_Horowitz, Joseph_Sumati}. The boundary conditions on the fields thus play a crucial role in determining the finiteness of the action. Considering spacetime as a cylindrical slab, it was shown in \cite{Ashtekar} that the action is finite (even off-shell) without any counterterm, i.e., $C=0$. This is also true in the presence of the supertranslations using the boundary conditions discussed above. This can be seen as follows. The eq. (\ref{S_modified}) can be re-expressed as 
\begin{equation}
    \label{actn2} S=\dfrac{1}{2\kappa} \int_{M} (\boldsymbol{d\Sigma}_{IJ} \wedge \boldsymbol{\omega}^{IJ} - \boldsymbol{\Sigma}_{IJ} \wedge \boldsymbol{\omega}^I_{\ K} \wedge \boldsymbol{\omega}^{KJ}).
\end{equation}
Since $\boldsymbol{d\Sigma}\sim \frac{1}{\rho^2}$, the leading fall-off of the integrand of eq. (\ref{actn2}) is $\mathcal{O}(\rho^{-4}).$ Further the leading behaviour of the volume element in bulk is $\sim\rho^3$, which on any Cauchy surface ($t=\text{const.}$) reduces to $\rho^2\sin{\theta}\ d\rho d\theta d\phi$. Therefore, the expression (\ref{actn2}) remains finite and no counterterm needs to be added. Hence, using Compère-Dehouck-Troessaert boundary conditions the action remains manifestly finite if the two
Cauchy surfaces are asymptotically time-translated with respect to each other \cite{Ashtekar, Corichi}.


\subsection{Symplectic structure}
\label{symp_structure}
To construct the covariant phase space \cite{Ashtekar_Bombelli_Reula, Crnkovic_Witten} one starts with the pre-symplectic potential that can be extracted from eq. (\ref{dS2}) as
\begin{equation}
    \Theta = \dfrac{1}{2\kappa} \int_{\partial M} \delta\boldsymbol{\Sigma}_{IJ} \wedge \boldsymbol{\omega}^{IJ},
\end{equation}
and the pre-symplectic current as
\begin{equation}
    \boldsymbol{J}[\delta_1,\delta_2] = \dfrac{1}{2\kappa} \big( \delta_1\boldsymbol{\Sigma}_{IJ} \wedge \delta_2\boldsymbol{\omega}^{IJ} - \delta_2\boldsymbol{\Sigma}_{IJ} \wedge \delta_1\boldsymbol{\omega}^{IJ} \big).
\end{equation}
One can vary the action a second time and verify that the pre-symplectic current is conserved, i.e., $\boldsymbol{dJ}=0$.

We consider the boundary as a cylindrical slab\footnote{The cylindrical slab contains $\Sigma_{1}$ and $\Sigma_2$ as $t=\text{constant}$ surfaces, which are asymptotically time-translated. One could also choose a boosted slab in which $\Sigma_{1}$ and $\Sigma_2$ are $\tau=\text{constant}$ surfaces.}, $\partial M = \Sigma_1 \cup \Sigma_2 \cup \tau_\infty$, where $\Sigma_{1,2}$ are Cauchy surfaces and $\tau_\infty$ is a cylindrical surface at spatial infinity. The symplectic structure is well-defined if there is no leakage through $\tau_\infty$, i.e., the flux of the symplectic current through $\tau_\infty$ must vanish, and the integral of the symplectic current on any Cauchy surface is finite. Let us check whether we have a well-defined symplectic structure.
\begin{equation}
    \int_M \boldsymbol{dJ} =\int_{\partial M} \boldsymbol{J} = \int_{\Sigma_1} \boldsymbol{J} - \int_{\Sigma_2} \boldsymbol{J} + \int_{\tau_\infty} \boldsymbol{J} = 0,
\end{equation}
where the negative sign in the second integral is due to the orientation of the Cauchy surface, i.e., it must be past-directed according to Stoke's theorem. Let us first consider the symplectic current at $\tau_\infty$:
\begin{multline}\label{J_Integrand}
    \lim_{\rho\to\infty} \big(\delta_{[1} \Sigma^{(1)}_{abIJ} \big)  \big(\delta_{2]} \omega_{c}^{(2)IJ} \big) \rho^{-3} \epsilon^{abc} \\= \lim_{\rho\to\infty} \epsilon_{IJKL} e_a^{(0)K} \big(\delta_{[1} e_b^{(1)L} \big) \big(\delta_{2]} \omega_{c}^{(2)IJ} \big) \rho^{-3} \epsilon^{abc}.
\end{multline}
Where $\epsilon^{abc}$ corresponds to the 3-volume form on $\tau_\infty$. Now, substitute $\boldsymbol{e}^{(1)I}$ and $\boldsymbol{\omega}^{(2)IJ}$ in the above expression and check in fact it vanishes.
\begin{multline}
    \therefore \lim_{\rho\to\infty} \epsilon_{IJKL} e_a^{(0)K} \big(\delta_{[1} e_b^{(1)L} \big) \big(\delta_{2]} \omega_{c}^{(2)IJ} \big) \rho^{-3} \epsilon^{abc} \\
    = \epsilon_{IJKL} e_a^{(0)K} \big(\rho_b \rho^L \delta_{[1}\sigma + \frac{1}{2}\delta_{[1}h_{bd}^{(1)} e^{(0)dL} \big) \\
    \cdot \big(2\rho\rho_c\partial_e\delta_{2]}\sigma e^{(0)e[J}\rho^{I]} - \delta_{2]}h^{(1)}_{ce} \rho^{[J} e^{(0)|e|I]} \\
    + \rho\partial_e\delta_{2]}h^{(1)}_{cf} e^{(0)e[J}e^{(0)|f|I]} \big) \rho^{-3} \epsilon^{abc} \\
    =\frac{1}{2}\rho \epsilon_{IJKL} e^{(0)K}_a e^{(0)fI} e^{(0)dL} (\delta_{[1}h^{(1)}_{bd}) (\partial^J\delta_{2]}h^{(1)}_{cf}) \rho^{-3} \epsilon^{abc} \\
    =0
\end{multline}
Where, $\rho_a\epsilon^{abc}=0$, and $e^{(0)\rho I}=0$ if $I \ne 1$. Next, one needs to evaluate eq. (\ref{J_Integrand}) on $\Sigma_1$ (or on $\Sigma_2$)  to check whether it is finite. The leading order term in eq. (\ref{J_Integrand}) falls off as $\mathcal{O}(\rho^{-3})$, and the volume element on any Cauchy surface is given by $\rho^2\sin{\theta} d\rho d\theta d\phi$. This indicates the integral can be logarithmically divergent \cite{Ashtekar}. However, expanding the leading order term as above and using the parity conditions described in (\ref{AFSTi0}) on $\sigma$ and $k_{ab}$ (or $\Psi$) one can see that the leading order term in eq. (\ref{J_Integrand}) on any Cauchy surface vanishes. As the remaining terms in (\ref{J_Integrand}) have a fall of $\mathcal{O}(\rho^{-4})$ or faster, the integral converges.

\begin{figure}
    \centering
    \includegraphics[width=0.7\linewidth]{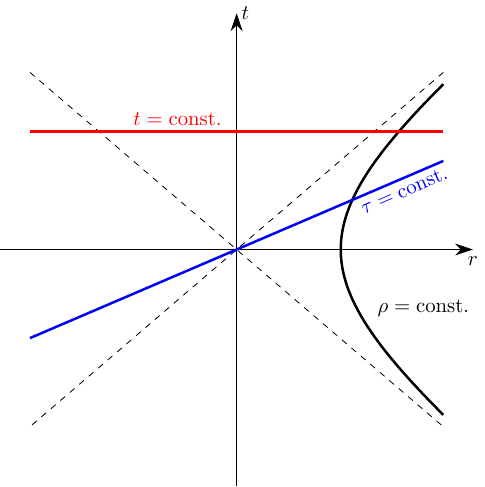}
    \caption{Visualization of $t=\text{constant}$ and $\tau=\text{constant}$ surfaces at $i^0$.}
    \label{fig1}
\end{figure}

Thus, we obtain a conserved pre-symplectic form:
\begin{equation}
    \Omega[\delta_1,\delta_2] = \dfrac{1}{2\kappa} \int_{\Sigma} \big( \delta_1\boldsymbol{\Sigma}_{IJ} \wedge \delta_2\boldsymbol{\omega}^{IJ} - \delta_2\boldsymbol{\Sigma}_{IJ} \wedge \delta_1\boldsymbol{\omega}^{IJ} \big),
    \label{Symp_2-form}
\end{equation}
where $\Sigma$ is any Cauchy surface. We choose $\Sigma$ to be $\tau=\text{constant}$ surface, see fig. (\ref{fig1}).

\vskip -0.3 cm
\subsection{Supertranslation charges}
\label{STCi0}
Let us consider the generator $\boldsymbol{\xi}$ on the manifold $M$ for the asymptotic symmetry transformations at spatial infinity. Using the generator $\boldsymbol{\xi}$ we define a Hamiltonian vector field $X_\xi$ on the phase space $\Gamma$:
\begin{equation}
    X_\xi = \int_M \mathcal{L}_\xi \phi^a \dfrac{\delta}{\delta\phi^a},
\end{equation}
where $\phi^a$ is a set of dynamical fields $(\boldsymbol{e},\ \boldsymbol{\omega})$. Now, we define the Hamiltonian charge as
\begin{equation}
    X_\xi \cdot \Omega(\delta_1,\delta_2) = -\delta H_\xi,
\end{equation}
where ``$\cdot$" represents the interior product.
\begin{align}
    \therefore \delta H_\xi &= -\Omega(\delta_\xi,\delta) \nonumber \\
    &= -\dfrac{1}{2\kappa} \int_{\Sigma} \big(\mathcal{L}_\xi \boldsymbol{\Sigma}_{IJ} \wedge \delta\boldsymbol{\omega}^{IJ} - \delta\boldsymbol{\Sigma}_{IJ} \wedge \mathcal{L}_\xi \boldsymbol{\omega}^{IJ} \big).
    \label{dH}
\end{align}
Now, we use the Cartan identities, equations of motion, and the linearized equations of motion for $\delta$ to obtain
\begin{equation}
    \delta H_\xi = \dfrac{1}{2\kappa} \oint_{S_
\infty} \big( (\boldsymbol{\xi}\cdot\boldsymbol{\omega}^{IJ})\delta\boldsymbol{\Sigma}_{IJ} - (\boldsymbol{\xi}\cdot\boldsymbol{\Sigma}_{IJ}) \wedge \delta\boldsymbol{\omega}^{IJ} \big),
\label{dH_xi}
\end{equation}
where $S_\infty$ is the boundary of the Cauchy surface $\Sigma$. The form of the above Hamiltonian charge is the same as obtained in \cite{Ashtekar}. The detailed calculation is given in appendix (\ref{A}). 

The first integrand vanishes in $\rho\to\infty$ limit as $\boldsymbol{\omega}^{IJ} \sim o(\rho^{-2})$, $\delta\boldsymbol{\Sigma}_{IJ}\sim o(\rho^{-1})$, and the volume element of $S_\infty$ grows as $\rho^2$. Therefore,
\begin{equation}
    \delta H_\xi = -\dfrac{1}{2\kappa} \oint_{S_
\infty} (\boldsymbol{\xi}\cdot\boldsymbol{\Sigma}_{IJ}) \wedge \delta\boldsymbol{\omega}^{IJ}.
\label{dH_xi_reduced}
\end{equation}
Now, we substitute $\boldsymbol{e}^I$ and $\boldsymbol{\omega}^{IJ}$ in the above expression to obtain
\begin{multline}
    \delta H_\xi = \dfrac{1}{2\kappa} \oint_{S_\infty} |e| \big(4\rho\rho_a\xi^a n^b \nabla_b\delta\sigma - 4n_a\xi^a\delta\sigma\\ 
    -\rho\xi^b\rho_b n^c \nabla^a\delta k_{ac} - \xi^a n^b \delta k_{ab}\\
    + \rho\xi^b\rho_b n^c\nabla_c \delta k + n_c\xi^c\delta k
    \big) d^2\theta,
    \label{dH_xi_1}
\end{multline}
where $|e|$ is the determinant of the tetrad $\boldsymbol{e}^I$, $n^a$ is the unit normal to the 2-sphere $S_\infty$ within the Hyperboloid $\mathcal{H}$. Setting $k_{ij}$ to be zero in the above expression yields eq. (4.5) of \cite{Ashtekar}. Extracting $\delta$ in eq. (\ref{dH_xi_1}) gives us
\begin{multline}
    H_\xi = \dfrac{1}{2\kappa} \oint_{S_\infty} \big(4\omega n^i D_i\sigma - 4n_iD^i\omega\sigma\\ 
    -\omega n^jD^i k_{ij} - D^i\omega n^j k_{ij}\\
    + \omega n^iD_i k + n_iD^i\omega k
    \big) d^2 S_0,
\end{multline}
where $\xi^a\rho_a=\omega$, $\xi^i=D^i\omega$, and $d^2S_0$ is the area element on the unit 2-sphere.

Now, we use the boundary conditions (\ref{BC1}), (\ref{BC2}). The above equation reduces to
\begin{multline}
    H_\xi = \dfrac{1}{2\kappa} \oint_{S_\infty} \big(4\omega n^i D_i\sigma - 4n_iD^i\omega\sigma - D^i\omega n^j k_{ij}
    \big) d^2 S_0.
\end{multline}
The last term can be converted to $D^i\omega n^j k_{ij}=D^i(\omega n^j k_{ij}) - \omega n^j D    ^i k_{ij}$ using the integration by parts, where the total derivative term vanishes due to the divergence theorem and the second term vanishes using the boundary conditions.
\begin{equation}
    H_\xi = \dfrac{2}{\kappa} \oint_{S_\infty} n_i \big(\omega D^i\sigma - \sigma D^i\omega
    \big) d^2 S_0,
\end{equation}
where $n_i$ is normal to the 2-sphere within $\mathcal{H}$. The above expression matches with the Compère-Dehouck expression for the supertranslation charges at $i^0$ \cite{Compere_Dehouck}.


\section{Asymptotic Structure at Future Timelike Infinity $i^+$}
\label{ASFTI}
In this section, we first define a suitable coordinate system to study $i^+$. We consider an asymptotically flat spacetime metric at $i^+$ derived in \cite{Sumanta_Virmani} and adopt the same boundary conditions as in sec. (\ref{AFSTi0}). 


\subsection{Flat spacetime at $i^+$}
In the spherical-polar coordinates, $i^+$ is approached as $t\to\infty$. This limit can not distinguish between the timelike trajectories and the null trajectories. One chooses rapidity as a coordinate:
\begin{equation}
    \rho = \tanh^{-1}\dfrac{r}{t},
\end{equation}
where the ranges of the coordinates $(\rho,\theta,\phi)$ are the same as the standard spherical-polar coordinates guarantee that we include only the massive particles. These new set of coordinates $\Phi^i=(\rho,\theta,\phi)$ parametrize the Euclidean $\text{AdS}_3$ hyperboloid $\mathcal{H}^+$ whose points represent velocities of the massive particles.

Since $i^+$ describes the timelike separated region, $t^2-r^2>0$. Thus, we consider proper time $\tau$ as the time coordinate:
\begin{equation}
    \tau = \sqrt{t^2-r^2},
\end{equation}
where
\begin{align}
    &t=\tau \cosh{\rho}, &r=\tau \sinh{\rho}.
\end{align}
Thus, the Minkowski metric in a new set of coordinates, $(\tau,\rho,\theta,\phi)$, takes the form
\begin{align}
    ds^2 &= -d\tau^2 + \tau^2 \left(d\rho^2 + \sinh^2\rho \ \gamma_{AB} dx^A dx^B \right) \nonumber \\
    &= -d\tau^2 + \tau^2 h^{(0)}_{ij}d\Phi^i d\Phi^j,
\end{align}
where $h^{(0)}_{ij}$ is the induced metric on $\mathcal{H}^+$.

\subsection{Asymptotically flat spacetime at $i^+$}
The asymptotically flat spacetime at $i^+$ can be put in the following form up to the second-order corrections \cite{Sumanta_Virmani}
\begin{multline}
    ds^2 = -\left(1+\dfrac{2\sigma(\Phi)}{\tau} + \dfrac{\sigma^2(\Phi)}{\tau^2} \right) d\tau^2 
    \\+ \tau^2 \left(h^{(0)}_{ij}(\Phi) + \dfrac{h^{(1)}_{ij}(\Phi)}{\tau} + \dfrac{h^{(2)}_{ij}(\Phi)}{\tau^2} + \mathcal{O}(\tau^{-3})_{ij} \right) d\Phi^i d\Phi^j,
    \label{metric i+}
\end{multline}
where $\sigma$ is a function of angular coordinates and $h^{(1)}_{ij}$, $h^{(2)}_{ij}$ are tensor fields on $\mathcal{H}^+$. The diffeomorphism
\begin{equation}
    \tau \to \tau - \omega(\Phi) + \mathcal{O}(\tau^{-1}),
\end{equation}
\begin{equation}
    \Phi^i \to \Phi^i + \dfrac{1}{\tau} D^i\omega(\Phi) + \mathcal{O}(\tau^{-2})
\end{equation}
preserves the above asymptotic form of the metric up to the first-order correction. Where $\omega$ is an arbitrary function of the angular coordinates. The above diffeomorphism is a general supertranslation at timelike infinity. Under the supertranslation, the function $\sigma$ and the field $k_{ij}$ transform as
\begin{align}
    &\sigma \to \sigma, \\
    &k_{ij} \to k_{ij} +2 D_i D_j \omega - 2\omega h^{(0)}_{ij}.
\end{align}
Where, $k_{ij} = h^{(1)}_{ij} + 2\sigma h^{(0)}_{ij}$.
We choose the same boundary conditions on $k_{ij}$ as in eqs. (\ref{BC1}) and (\ref{BC2}). The class of supertranslations which preserves these boundary conditions is then given by
\begin{equation}
    \left(D^2 - 3\right)\omega = 0,
\end{equation}
where $D^2=D^iD_i$ is the Laplacian on $\mathcal{H}^+$.
As already seen in sec. (\ref{STCi0}), the first-order terms in the metric are sufficient to compute the charges. Thus, we truncate the metric (\ref{metric i+}) to the first-order terms:
\begin{equation}
    ds^2 = -\left(1+\dfrac{2\sigma}{\tau} \right) d\tau^2 + \tau^2 \left(h^{(0)}_{ij} + \dfrac{1}{\tau} h^{(1)}_{ij} \right) d\Phi^i d\Phi^j.
\end{equation}
The asymptotic expansions of the tetrad and the connection one-form at $i^+$ are,
\begin{equation}
    \boldsymbol{e}^I(\tau,\Phi) = \boldsymbol{e}^{(0)I}(\Phi) + \dfrac{1}{\tau} \boldsymbol{e}^{(1)I}(\Phi) + ...
\end{equation}
\begin{multline}
    \boldsymbol{\omega}^{IJ}(\tau,\Phi) = \boldsymbol{\omega}^{(0)IJ}(\tau,\Phi) + \dfrac{1}{\tau} \boldsymbol{\omega}^{(1)IJ}(\tau,\Phi) \\
    + \dfrac{1}{\tau^2} \boldsymbol{\omega}^{(2)IJ}(\tau,\Phi) + ....
\end{multline}
As in sec. (\ref{First-Order}), one can compute the coefficients in the asymptotic expansion of $\boldsymbol{e}^I$ and $\boldsymbol{\omega}^{IJ}$, which are as follows:
\begin{align}
    e^{(1)I}_{a} &= -\sigma \tau_a \tau^I + \dfrac{1}{2}h^{(1)}_{ab} e^{(0)bI} \nonumber\\
    &=-\sigma(2\tau_a\tau^I + e_a^{(0)I}) + \dfrac{1}{2}k_{ab}e^{(0)bI},
\end{align}
\begin{equation}
    \omega^{(0)IJ}_a = 0 = \omega^{(1)IJ}_a,
\end{equation}
\begin{multline}
    \omega^{(2)IJ}_a = -4\tau\tau_a\tau^{[I}\partial^{J]}\sigma - 2\tau e^{(0)[I}_a\partial^{J]}\sigma + \tau e^{(0)b[I}\partial^{J]}k_{ab} \\
    + 2\sigma e^{(0)[I}_a\tau^{J]} + k_{ab}e^{(0)b[I}\tau^{J]}.
\end{multline}
Where, $\tau_a=-\partial_a\tau$ is a normal to $\mathcal{H}^+$, and $\tau^I=\tau_a e^{(0)aI}$.


\subsection{Supertranslation charges at $i^+$}
\begin{figure}
    \centering
    \includegraphics[width=0.7\linewidth]{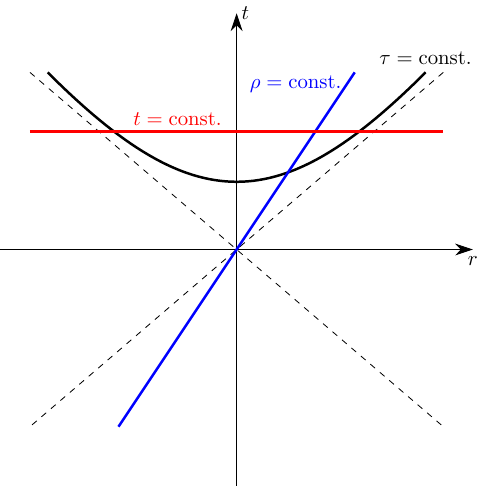}
    \caption{Visualization of $t=\text{constant}$ and $\tau=\text{constant}$ surfaces at $i^+$.}
    \label{fig2}
\end{figure}

We define a pre-symplectic form at $i^+$ following  eq. (\ref{Symp_2-form}). The Cauchy surface $\Sigma$ on which it is defined is taken to be a union of the hypersurface $\mathcal{H}^+$ at $\tau\to\infty$, and $\mathscr{I}^+$, i.e. $\Sigma=i^+\cup \mathscr{I}^+$. This is because a $\tau=\text{constant}$ hypersurface fails to be a Cauchy surface near $i^+$, see fig. (\ref{fig2}). 

We first evaluate the contribution to pre-symplectic current from $i^+$. This is given by
\begin{equation}
    \lim_{\tau\to\infty} \big(\delta_{[1} \Sigma^{(1)}_{abIJ} \big)  \big(\delta_{2]} \omega_{c}^{(2)IJ} \big) \tau^{-3} \epsilon^{abc},
\end{equation}
where $\epsilon^{abc}$ is a metric compatible 3-form. The volume element on a $\tau=\text{constant}$ surface grows as $\tau^3$, thus the above integrand seems to be finite. But, we show that it in fact vanishes.
\begin{multline}
   \lim_{\tau\to\infty} \epsilon_{IJKL} e_a^{(0)K} \big(\delta_{[1} e_b^{(1)L} \big) \big(\delta_{2]} \omega_{c}^{(2)IJ} \big) \tau^{-3} \epsilon^{abc} \\
    = \epsilon_{IJKL} e_a^{(0)K} \big(-\tau_b \tau^L \delta_{[1}\sigma + \frac{1}{2}\delta_{[1}h_{bd}^{(1)} e^{(0)dL} \big) \\
    \cdot \big(-2\tau\tau_c\partial_e\delta_{2]}\sigma e^{(0)e[J}\tau^{I]} + \delta_{2]}h^{(1)}_{ce} \tau^{[J} e^{(0)|e|I]} \\
    + \tau\partial_e\delta_{2]}h^{(1)}_{cf} e^{(0)e[J}e^{(0)|f|I]} \big) \tau^{-3} \epsilon^{abc} \\
    =\frac{1}{2}\tau \epsilon_{IJKL} e^{(0)K}_a e^{(0)fI} e^{(0)dL} (\delta_{[1}h^{(1)}_{bd}) (\partial^J\delta_{2]}h^{(1)}_{cf}) \tau^{-3} \epsilon^{abc}\\
    = 0,
\end{multline}
where in the third line we have used the identity $\tau_a \epsilon^{abc}=0$, and in the last line $e^{(0)K}_a \epsilon^{abc}=0$; $K=0$. Thus, there is no contribution to charges from $i^+$, which has been shown in \cite{Sumanta_Virmani} using the metric-based approach. The vanishing of the pre-symplectic current ($\delta H_{\xi}$) enables us to define a charge on any 2-dimensional topological sphere surrounded by ``sources" at $i^+$.

Next, we take the limit $\rho\to\infty$ to determine the contribution from $\mathscr{I}^+$ to the charges. In this limit, the $\tau=\text{constant}$ surface at $i^+$ reaches $\mathscr{I}^+_+$ the future boundary of $\mathscr{I}^+$. This allows us to define (local) conserved charges for supertranslations at timelike infinity. 

It is easy to see that the expression for the charges formally remains the same as eq. (\ref{dH_xi_reduced}). Now we substitute the asymptotic fields in eq. (\ref{dH_xi_reduced}) to obtain
\begin{multline}
    H_\xi = \dfrac{1}{2\kappa} \oint_{S_
\infty} (4\tau\tau_a\xi^a\rho_b\nabla^b\sigma - 4\rho_b\xi^b \sigma + \tau\tau_b\xi^b\rho_c\nabla^a k_{a}^c\\
-\tau\tau_b\xi^b\rho_c\nabla^c k + \rho_b\xi^a k_{a}^b - \rho_b\xi^b k
) d^2S_0.
\label{i+chg}\end{multline}
Where, $\tau_a\xi^a=\omega$, $\rho_a\xi^a = \rho_i D^i\omega$. Using the boundary conditions on $k_{ij}$ and integration by parts as in sec. (\ref{STCi0}), the above expression is reduced to
\begin{equation}
    H_\xi = \dfrac{2}{\kappa} \oint_{S_
\infty} \rho_i(\omega D^i\sigma - \sigma D^i\omega) d^2S_0,
\label{Chgi+m}\end{equation}
where $\rho_i$ is normal to the 2-sphere within $\mathcal{H}^+$. This is the expression for the supertranslation charges (``future charges") at $i^+$ proposed in \cite{Sumanta_Virmani}.

Since all massive bodies reach in the (future) time like infinity and they can not always be represented by weak fields, the interpretation of charges and their conservations becomes subtle. 
 
For a system composed of individually bound objects but gravitationally unbound relative to each other, timelike infinity can be taken to be $\mathcal{H}^+$ minus one point each for the individually bound objects \cite{Porrill_1982}. These points act as sources for the charges \cite{Compere, Sumanta_Virmani}. The charge expression (\ref{i+chg}) or (\ref{Chgi+m}) can be understood as a Gauss-law in the presence of those point charges. In fact, eq. (\ref{Chgi+m}) can be rearranged in a form including a current: $\int_{S_
\infty}d^2S_0J_i\rho^i$ to show $D^ij_i=D^i(\omega D_i\sigma - \sigma D_i\omega)=0$ upon using the Einstein's equations. The charge integral then becomes independent of the particular topological 2-sphere chosen. This is the sense in which the charge is conserved.  In the presence of massive bodies, the integral can be performed considering a 2-sphere surrounding an individual object. When the integral is pushed towards $\rho \to \infty$ the corresponding expression contains the linear sum of the charges due to each such body, and it serves as the total charge for the system (see \cite{Compere}, and \cite{Porrill_1982} for more details).


\section{Discussions}
\label{Discussions}
In this work, we have examined the supertranslations at spatial infinity and future timelike infinity utilizing the first-order framework. We considered the Beig-Schmidt expanded forms (up to the first order terms) of the asymptotically flat metric at both the infinities. We imposed suitable boundary conditions that allowed a specific class of supertranslations at these infinities \cite{Compere, Sumanta_Virmani}. We then obtained the desired form of the co-tetrad and the Lorentz connection and used them to show that the covariant phase space at $i^0$ is well-defined. Further, we showed that a finite conserved pre-symplectic form exists. Next, constructing the Hamiltonian vector field corresponding to the supertranslations, we obtained the corresponding charges that match with the known results at spatial infinity \cite{Compere_Dehouck, Troessaert_2018}. For timelike infinity, we followed a similar approach and proposed conserved supertranslation charges  \cite{Sumanta_Virmani, Compere}. This work would be useful to address the issue of matching the supertranslation charges across the spatial infinity in first-order framework. However, this needs a first-order framework at the null infinities. Further, inclusion of future horizons would be important to understand the conservation at $i^+$. It will also be interesting to see how the first-order framework can accommodate various extensions of the BMS group at different boundaries of AF spacetimes.

In this study, we considered only the first order terms of the Beig-Schmidt metric and switched off the logarithmic translations by imposing boundary conditions. Extending the present analysis to include the log terms and higher order terms would be interesting for several reasons. One may wish to compare the charges for logarithmic translations obtained through a first-order formalism with the ones reported in the metric-based approach \cite{Compere}. The second order terms would also be necessary if one wishes to compute the Lorentz charges, which we have ignored in this note. The finite Lorentz charges were deduced in \cite{Ashtekar} in the absence of supertranslations. To obtain a finite charge for the relativistic angular momentum, it was necessary to expand the connection $\boldsymbol{\omega}$ up to the third order $o(\rho^{-3})$. This will be the case in the presence of supertranslations also. We wish to address these issues in a future communication. 

 It is known that in the presence of supertranslations, the relativistic angular momentum (Lorentz generators) suffers from ambiguities. This is apparent from the fact that the Poisson bracket between the supertranslations and Lorentz charges does not vanish. Therefore the angular momentum does not remain invariant under the supertranslations. This issue has recently been addressed and resolved by Chen et al.\cite{Chen_et_al_2021} at null infinity and by 
 Fuentealba et al. \cite{Fuentealba_et_al_2024, Fuentealba_Henneaux_Tross} at spatial infinity. These developments inspired people to define \textit{supertranslation invariant Lorentz charges}. Since these charges have also been defined using the Beig-Schmidt framework \cite{Sumanta_Virmani_2024}, we believe the same can be achieved in the first order framework as well. We wish to study this aspect in the future.  
 
\begin{acknowledgments}
It is a pleasure to thank Amitabh Virmani for several discussions at various stages of this work and for his comments on a preliminary version of the draft. Research of S.B. is supported by SERB-DST through the MATRICS grant MTR/2022/000170. 
\end{acknowledgments}

\appendix
\section{}
\label{A}
Let us consider the contracted pre-symplectic form (\ref{dH})
\begin{equation}
    \delta H_\xi = -\dfrac{1}{2\kappa} \int_{\Sigma} \big(\mathcal{L}_\xi \boldsymbol{\Sigma}_{IJ} \wedge \delta\boldsymbol{\omega}^{IJ} - \delta\boldsymbol{\Sigma}_{IJ} \wedge \mathcal{L}_\xi \boldsymbol{\omega}^{IJ} \big).
\end{equation}
We use the Cartan identities:
\begin{align}
    &\mathcal{L}_\xi \boldsymbol{\Sigma}_{IJ} = (\boldsymbol{d\xi}\cdot + \boldsymbol{\xi}\cdot \boldsymbol{d}) \boldsymbol{\Sigma}_{IJ}, \\
    &\mathcal{L}_\xi \boldsymbol{\omega}^{IJ} = (\boldsymbol{d\xi}\cdot + \boldsymbol{\xi}\cdot \boldsymbol{d}) \boldsymbol{\omega}^{IJ}.
\end{align}
Therefore,
\begin{multline}
    \delta H_\xi = -\dfrac{1}{2\kappa} \int_{\Sigma} \big(\boldsymbol{d}(\boldsymbol{\xi}\cdot\boldsymbol{\Sigma}_{IJ})\wedge\delta\boldsymbol{\omega}^{IJ} + \boldsymbol{\xi}\cdot \boldsymbol{d\Sigma}_{IJ}\wedge\delta\boldsymbol{\omega}^{IJ}\\
    -\delta\boldsymbol{\Sigma}_{IJ}\wedge \boldsymbol{d}(\boldsymbol{\xi}\cdot\boldsymbol{\omega}^{IJ}) - \delta\boldsymbol{\Sigma}_{IJ}\wedge\boldsymbol{\xi}\cdot \boldsymbol{d\omega}^{IJ} \big).
    \label{A4}
\end{multline}
Now, we separate the total derivative terms by substituting
\begin{equation}
    \boldsymbol{d}(\boldsymbol{\xi}\cdot\boldsymbol{\Sigma}_{IJ})\wedge\delta\boldsymbol{\omega}^{IJ} = \boldsymbol{d}(\boldsymbol{\xi}\cdot\boldsymbol{\Sigma}_{IJ} \wedge \delta\boldsymbol{\omega}^{IJ}) + \boldsymbol{\xi}\cdot\boldsymbol{\Sigma}_{IJ} \wedge \delta \boldsymbol{d\omega}^{IJ}
\end{equation}
and 
\begin{equation}
    \delta\boldsymbol{\Sigma}_{IJ}\wedge \boldsymbol{d}(\boldsymbol{\xi}\cdot\boldsymbol{\omega}^{IJ}) = \boldsymbol{d}(\delta\boldsymbol{\Sigma}_{IJ}\wedge \boldsymbol{\xi}\cdot\boldsymbol{\omega}^{IJ}) - \delta \boldsymbol{d\Sigma}_{IJ}\wedge \boldsymbol{\xi}\cdot\boldsymbol{\omega}^{IJ}.
\end{equation}
Thus, eq. (\ref{A4}) reduces to
\begin{multline}
    \delta H_\xi = -\dfrac{1}{2\kappa} \oint_{S_\infty} (\boldsymbol{\xi}\cdot\boldsymbol{\Sigma}_{IJ}\wedge\delta\boldsymbol{\omega}^{IJ} - \delta\boldsymbol{\Sigma}_{IJ}\wedge\boldsymbol{\xi}\cdot\boldsymbol{\omega}^{IJ})\\
    +\dfrac{1}{2\kappa}\int_{\Sigma} (-\boldsymbol{\xi}\cdot\boldsymbol{\Sigma}_{IJ} \wedge \delta \boldsymbol{d\omega}^{IJ} - \boldsymbol{\xi}\cdot \boldsymbol{d\Sigma}_{IJ}\wedge \delta\boldsymbol{\omega}^{IJ}\\ - \delta \boldsymbol{d\Sigma}_{IJ}\wedge\boldsymbol{\xi}\cdot\boldsymbol{\omega}^{IJ} + \delta\boldsymbol{\Sigma}_{IJ}\wedge\boldsymbol{\xi}\cdot \boldsymbol{d\omega}^{IJ}).
    \label{A7}
\end{multline}
Now, using the equations of motion and the linearized equations of motion, we show that the integral on $\Sigma$ in the above expression in fact vanishes. The equation of motion is obtained by varying the action, i.e., from eq. (\ref{dS2}). Variation of the action w.r.t. $\boldsymbol{e}^I$ and $\boldsymbol{\omega}^{IJ}$ gives us the following equations of motion respectively.
\begin{equation}
    \epsilon_{IJKL} \boldsymbol{F}^{IJ} \wedge \boldsymbol{e}^K = 0,
    \label{eom}
\end{equation}
\vskip -0.2cm
\begin{equation}
    \boldsymbol{\mathcal{D}\Sigma}_{IJ} = \boldsymbol{d\Sigma}_{IJ} + \boldsymbol{\omega}_{I}^{\ K}\wedge\boldsymbol{\Sigma}_{KJ} + \boldsymbol{\omega}_{J}^{\ K}\wedge\boldsymbol{\Sigma}_{IK} = 0.
\end{equation}
Therefore,
\begin{equation}
    \boldsymbol{d\Sigma}_{IJ} = -\boldsymbol{\omega}_{I}^{\ K}\wedge\boldsymbol{\Sigma}_{KJ} - \boldsymbol{\omega}_{J}^{\ K}\wedge\boldsymbol{\Sigma}_{IK}.
    \label{d_Sigma}
\end{equation}
The curvature two-form is given by
\begin{equation}
    \boldsymbol{F}^{IJ} = \boldsymbol{d\omega}^{IJ} + \boldsymbol{\omega}^{I}_{\ K}\wedge\boldsymbol{\omega}^{KJ},
\end{equation}
\vskip -0.2cm
\begin{equation}
    \therefore \boldsymbol{d\omega}^{IJ} = \boldsymbol{F}^{IJ} - \boldsymbol{\omega}^{I}_{\ K}\wedge\boldsymbol{\omega}^{KJ}.
    \label{d_omega}
\end{equation}
Now, we substitute eqs. (\ref{d_omega}) and (\ref{d_Sigma}) into the integrand:
\begin{multline}
    \delta\boldsymbol{\Sigma}_{IJ}\wedge\boldsymbol{\xi}\cdot \boldsymbol{d\omega}^{IJ} - \delta \boldsymbol{d\Sigma}_{IJ}\wedge\boldsymbol{\xi}\cdot\boldsymbol{\omega}^{IJ}\\
    -\boldsymbol{\xi}\cdot\boldsymbol{\Sigma}_{IJ} \wedge \delta \boldsymbol{d\omega}^{IJ} - \boldsymbol{\xi}\cdot \boldsymbol{d\Sigma}_{IJ}\wedge \delta\boldsymbol{\omega}^{IJ} \\
    =\delta\boldsymbol{\Sigma}_{IJ}\wedge\boldsymbol{\xi}\cdot \boldsymbol{F}^{IJ} - \boldsymbol{\xi}\cdot\boldsymbol{\Sigma}_{IJ}\wedge\delta \boldsymbol{F}^{IJ}. \ \ \ \  \ \ \ \ \ \ \ \ \ \
\end{multline}
Now, let us take the variation in the equation of motion (\ref{eom}):
\begin{equation}
    \epsilon_{IJKL} \delta \boldsymbol{F}^{IJ}\wedge \boldsymbol{e}^K + \epsilon_{IJKL} \boldsymbol{F}^{IJ}\wedge \delta \boldsymbol{e}^K = 0,
\end{equation}
which is the linearized equation of motion.
\vskip -0.5cm
\begin{equation}
    \therefore \epsilon_{IJKL} \delta \boldsymbol{F}^{IJ}\wedge \boldsymbol{e}^K = -\epsilon_{IJKL} \boldsymbol{F}^{IJ}\wedge \delta \boldsymbol{e}^K.
    \label{leom}
\end{equation}
\vskip -0.2cm
Thus,
\begin{multline}
    \delta\boldsymbol{\Sigma}_{IJ}\wedge\boldsymbol{\xi}\cdot \boldsymbol{F}^{IJ} - \boldsymbol{\xi}\cdot\boldsymbol{\Sigma}_{IJ}\wedge\delta \boldsymbol{F}^{IJ} \\
    = \epsilon_{IJKL}\boldsymbol{e}^K\wedge\delta \boldsymbol{e}^L\wedge\boldsymbol{\xi}\cdot \boldsymbol{F}^{IJ} - \epsilon_{IJKL}\boldsymbol{\xi}\cdot \boldsymbol{e}^L \boldsymbol{F}^{IJ}\wedge\delta \boldsymbol{e}^K.
\end{multline}
In the second line the linearized equation of motion (\ref{leom}) is used. The above expression further simplifies to
\begin{align}
    &\epsilon_{IJKL}\boldsymbol{e}^K\wedge\delta \boldsymbol{e}^L\wedge\boldsymbol{\xi}\cdot \boldsymbol{F}^{IJ} - \epsilon_{IJKL}\boldsymbol{\xi}\cdot \boldsymbol{e}^L \boldsymbol{F}^{IJ}\wedge\delta \boldsymbol{e}^K \nonumber \\
    =&\delta \boldsymbol{e}^L \wedge \big(\epsilon_{IJKL}(\boldsymbol{\xi}\cdot \boldsymbol{F}^{IJ}\wedge \boldsymbol{e}^K + \boldsymbol{F}^{IJ}\boldsymbol{\xi}\cdot \boldsymbol{e}^K)\big) \nonumber \\
    =&\delta \boldsymbol{e}^L \wedge \boldsymbol{\xi}\cdot(\epsilon_{IJKL}\boldsymbol{F}^{IJ}\wedge \boldsymbol{e}^K) \nonumber \\
    =&0 \nonumber.
\end{align}
In the last line, the equation of motion (\ref{eom}) is used. Hence, the integral on $\Sigma$ in eq. (\ref{A7}) vanishes, and we obtain
\begin{equation}
    \delta H_\xi = -\dfrac{1}{2\kappa} \oint_{S_\infty} (\boldsymbol{\xi}\cdot\boldsymbol{\Sigma}_{IJ}\wedge\delta\boldsymbol{\omega}^{IJ} - \delta\boldsymbol{\Sigma}_{IJ}\wedge\boldsymbol{\xi}\cdot\boldsymbol{\omega}^{IJ}).
\end{equation}


\bibliography{references}

\end{document}